\advance\topmargin by 2.0in
\documentclass[aps,prl,twocolumn]{revtex4}
\usepackage{graphicx}
\usepackage{epsfig}

\begin{document}

\title{Exactly solvable case of a one-dimensional Bose-Fermi mixture}

\author{Adilet Imambekov and Eugene Demler}
\affiliation{Department of Physics, Harvard University, Cambridge
MA 02138}

\date{\today}

\begin{abstract}
We consider a one dimensional interacting bose-fermi mixture with
equal masses of bosons and fermions, and with equal and repulsive
interactions between bose-fermi and bose-bose particles. Such a
system can be realized in experiments with ultracold boson and
fermion isotopes in optical lattices. We
use the Bethe-ansatz technique to find the ground state energy at
zero temperature for any value of interaction strength and density
ratio between bosons and fermions. We prove that the mixture is 
always stable against demixing. Combining exact solution with
the local density approximation we calculate density profiles and
collective oscillation modes in a harmonic trap. 
In the strongly interating regime we use exact wavefunctions to calculate
correlation functions for bosons and fermions under periodic boundary conditions.

\end{abstract}

\maketitle Recent developments in the cooling and trapping of
atomic gases open exciting opportunities for experimental studies
of interacting systems under well-controlled conditions. Using
Feshbach resonances\cite{Feshbach} and/or optical
lattices\cite{Jaksch98,Bloch}  it is possible to reach strongly
interacting regimes, where correlations between atoms play a
crucial role. The effect of interactions is most prominent for low
dimensional systems, and recent experimental
realization\cite{Weiss,Paredes} of a strongly interacting
Tonks-Girardeau (TG) gas of bosons opens new perspectives in
experimental studies of strongly interacting systems in
1D\cite{Moritz1dmolecules}. A different line of research, which
has attracted considerable attention both
theoretically\cite{bftheory} and
experimentally\cite{bfexp,fermipressure}, is the study of atomic
mixtures. 1D Fermi gases have been theoretically well
investigated\cite{Giamarchi}, since they correspond to interacting
electrons in solid state quasi-1D  systems. On the other hand, 1D
bose-fermi(BF) mixtures did not attract equal attention until
recently\cite{Das, CazalillaHo, CDW, Mathey}, when it became
possible to realize such systems in experiments with cold atoms.
Several properties of such systems have been investigated so far,
including phase separation, fermion pairing and charge density
wave (CDW) formation. However, most of these investigations relied
on the  mean-field approximations or the Luttinger liquid (LL)
 formalism. The mean-field approximation is known to be unreliable in 1D, and  the LL approach does not allow the calculation of LL parameters
 in the strongly interacting regime. In this paper we study BF mixtures in the regime where an exact Bethe-ansatz solution is available. We use the exact solution to
 calculate the ground state energy and investigate phase separation and collective modes. 
 In the strongly interacting regime we derive an analytical determinant formula, which allows to calculate correlation functions at all distances numerically for a polynomial time in the system size. The exactly solvable case considered in this paper is relevant to current experiments, and can be used as a benchmark to check the validity of different approaches.

   A 1D interacting BF mixture is described by the Hamiltonian
  \begin{eqnarray}
H=\int_0^L dx (\frac{\hbar^2}{2m_b}\partial_x \Psi_b^\dagger\partial_x \Psi_b+\frac{\hbar^2}{2m_f}\partial_x \Psi_f^\dagger\partial_x \Psi_f) + \nonumber  \\
\int_0^L dx(\frac12g_{bb}\Psi_b^\dagger \Psi_b^\dagger\Psi_b\Psi_b
+ g_{bf} \Psi_b^\dagger \Psi_f^\dagger\Psi_f\Psi_b ).
\label{initialhamiltoniansc}
\end{eqnarray}
Here, $\Psi_b, \Psi_f$ are boson and fermion operators, $m_b, m_f$
are the masses, and $g_{bb}, g_{bf}$ are bose-bose  and bose-fermi
interaction strengths.
 An array of such one dimensional tubes can be realized using a strong optical lattice in $y,z$ directions\cite{Weiss,Paredes,Moritz}.
 Alternatively, one can create a single BEC in a 1D box potential using the techniques of Ref. \cite{BECinbox}.
 The model (\ref{initialhamiltoniansc}) is exactly solvable, when
\begin{eqnarray}
m_f=m_b\equiv m, \,  g_{bb}=g_{bf}\equiv g.\label{intercond}
\end{eqnarray}
The first condition is approximately satisfied for isotopes
of atoms. Some of the promising candidates are
$^{171}Yb+^{172}Yb$\cite{Yb}, $^{39(41)}K-^{40}K,$\cite{Cote} and
$^{86(84)}Rb-^{87(85)}Rb$\cite{Rb}. Alternatively, if one uses  an
additional optical lattice along  the $x$ direction with filling
factors smaller than one, then  (\ref{initialhamiltoniansc}) is an
effective Hamiltonian describing this system with the masses
determined by the tunneling\cite{Paredes}. Thus one can simulate
(\ref{initialhamiltoniansc}) using already available degenerate
mixtures\cite{bfexp}. We point out that to satisfy the  second condition in 
(\ref{intercond}), it is sufficient to have equal (positive) signs
for the two scattering lengths, but not necessarily their
magnitudes. Well away from confinement induced
resonances\cite{Olshanii98}, 1D interactions are given by $
g_{bb}=2\hbar \omega_{b\perp} a_{bb}, g_{bf}=2\hbar
\sqrt{\omega_{b\perp}\omega_{f\perp}}a_{bf}, $ where
$\omega_{b\perp},\omega_{f\perp}$ are radial confinement
frequencies, and $a_{bb},a_{bf}$ are 3D scattering lengths. For a
fixed value of $a_{bb}/a_{bf},$ one can always choose the detuning
of the  optical lattice laser frequencies in such a way that $g_{bb}=g_{bf}.$
Additionally, all interactions can be tuned using available BF Feshbach resonances\cite{Feshbach}.

\begin{figure}
\psfig{file=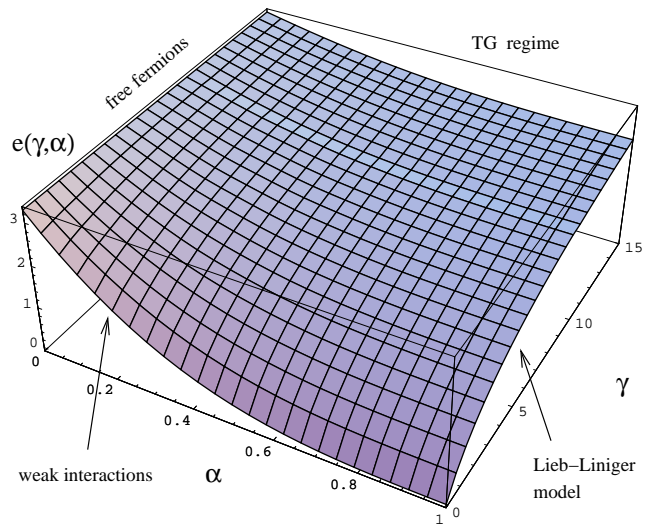} \caption{ \label{egammamn} Energy of
the ground state is given by $E=e(\gamma,\alpha) \hbar^2 N^3 /(2 m
L^2),$ where $\gamma=m g /(\hbar^2 n),$ and $\alpha=N_b/N$ is the
boson fraction. When $\alpha=0,$ system is purely fermionic, and
the energy doesn't depend on interactions. When $\alpha=1,$ the
system is purely bosonic, and numerically obtained energy
coincides with the result of \cite{LL}. If $\gamma=0,$ bosons and
fermions don't interact, and
$e(\gamma,\alpha)=(\pi^2/3)(1-\alpha)^3.$}
\end{figure}

Model (\ref{initialhamiltoniansc}) under conditions
 (\ref{intercond}) 
has been
considered in the literature before\cite{Lai}, but
its properties have not been investigated in detail. Similar to
the case of bosons \cite{LL}, a dimensionless parameter which
controls the strength of interactions is $\gamma=m g /(\hbar^2
n),$ with $n$ being the total density of particles. $\gamma \gg1$
corresponds to the strongly interacting mixture. The ground state
energy of the model (\ref{initialhamiltoniansc}), (\ref{intercond}) with periodic boundary
conditions is related to the solution of a system of integral
equations
\begin{eqnarray}
2 \pi \rho(k)=1 +\int_{-B}^{B}\frac{4c \sigma(\Lambda) d\Lambda }{c^2+4(\Lambda-k)^2} ,\label{bfequations31}\\
2 \pi \sigma(\Lambda)=\int_{-Q}^{Q}\frac{4c \rho(\omega) d\omega }{c^2+4(\Lambda-\omega)^2}.
\label{bfequations32}
\end{eqnarray}
Here $c=m g /\hbar^2,$  and $Q,B$ are related to densities via
\begin{eqnarray}
N/L=\int_{-Q}^{Q}\rho(k) dk, \, N_b/L=\int_{-B}^{B}\sigma(\Lambda) d\Lambda, \label{normm}
\end{eqnarray}
where $N_b$ is the number of bosons, $N$ is the  total number of particles, and $L$ is the length of the system.
The ground state energy can be expressed as
\begin{eqnarray}
E/L=\frac{\hbar^2}{2m}\int_{-Q}^{Q} k^2 \rho(k) dk.\label{energy}
\end{eqnarray}
One can solve eqs.  (\ref{bfequations31}-\ref{normm}) numerically  and  calculate the energy as a function of $\gamma$ and relative boson
density $\alpha=N_b/N.$ After rescaling $E=e(\gamma,\alpha) \hbar^2N^3/(2 m L^2), $ where $e(\gamma,\alpha)$ is shown in
fig \ref{egammamn}. In the limit $\gamma\gg1,$ one can derive the large $\gamma$ expansion of the energy
$e(\gamma,\alpha)=(\pi\gamma)^2/3(\gamma+2(\alpha + \sin{(\pi \alpha})/\pi ))^2 .$
Chemical potentials for bosons and fermions are given by
$
\mu_b=\partial E/\partial N_b, \mu_f=\partial E/\partial N_f,
$
where $N_f=N-N_b.$
Using the exact solution, one can analyze demixing
instabilities\cite{CazalillaHo, Das} for repulsive BF mixtures. In
the absence of an external potential the BF mixture is stable if the
compressibility matrix $\partial\mu_{b(f)}/\partial{n_{b(f)}} $ is
positively defined(here, $n_{b(f)}$ is a boson(fermion) density).
In our system we verified numerically that this condition is
satisfied for all $\alpha$ and $\gamma.$ This  proves that  the BF mixture under
conditions  (\ref{intercond}) is stable against
demixing for any strength of interaction. This is in contrast to
mean field approximation which predicts demixing for strong enough
interactions\cite{Das}.
Although an exact solution is available only  under conditions
(\ref{intercond}), small deviations from these
should not dramatically change the energy $e(\gamma,\alpha).$
Therefore, we expect the 1D mixtures to remain stable to demixing
in the vicinity of the integrable line
(\ref{intercond}) for any interaction strength.

\begin{figure}
\psfig{file=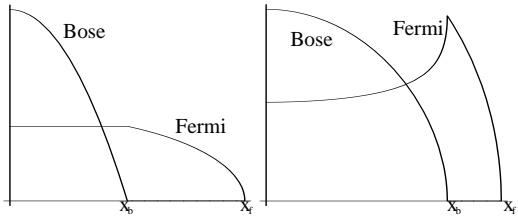} \caption{ \label{lda} Densities of Bose and
Fermi gases in weakly interacting(left , $\gamma_0=0.18$) and
strongly interacting($\gamma_0 \gg 1$) regimes. The overall number
of bosons equals the number of fermions.}
\end{figure}

Using numerically obtained $e(\gamma,\alpha)$ and local density
approximation(LDA), one can investigate the behavior of the system
in an external field.
 Local density approximation
is expected to be valid for slowly varying external potential. We
will assume that the confinement frequency for bosons and fermions
is the same. 
Within LDA, density distributions in the region of
coexistence are governed by a set of equations $
\mu_f(n_b^0(x),n_f^0(x))+ \frac{m \omega^2_0 x^2}2=\mu_f(n_b^0(0),n_f^0(0)) $ and $
\mu_b(n_b^0(x),n_f^0(x)) +\frac{m \omega^2_0 x^2}2=\mu_f(n_b^0(0),n_f^0(0)). $ It turns
out that this set of equations cannot be satisfied for the whole
cloud, and consequently a phase separation  occurs in the  presence
 of such external potential.
The BF mixture is present in the central part, but the outer
sections consist of Fermi gas only. In the weakly interacting
limit, this can be interpreted as an effect of the Fermi
pressure\cite{fermipressure}. 
While bosons can condense to the center of the trap, Pauli principle pushes fermions apart. 
As interactions get stronger, the relative distribution of bosons and
fermions changes, and fig \ref{lda} contrasts the limits of strong
and weak interactions. 
For strong interactions, the fermi density
shows strong non-monotonous behavior.

Recent experiments\cite{Moritz} demonstrated that collective
oscillations of 1D gases provide useful information about
interactions in the system. Within the region of coexistence of
bosons and fermions, such oscillations can be described by four
hydrodynamic equations\cite{Menotti}
\begin{eqnarray}
\frac{\partial}{\partial t}\delta n_{b(f)}+\frac{\partial}{\partial x}(n_{b(f)} v_{b(f)})=0,\label{dndt}\\
m\frac{\partial}{\partial t}\delta v_{b(f)}+\frac{\partial}{\partial x}(\mu_{b(f)}+V_{ext, b(f)}+\frac12 m v_{b(f)}^2)=0.\label{dvdt}
\end{eqnarray}
In certain cases, analytical solutions of hydrodynamic eqs. are
available\cite{Stringari, Menotti} and provide the frequencies of
collective modes. When an analytic solution is not available, the
"sum rule'' approach has been used\cite{Stringari, Menotti,Astrakharchik,bfsumrule}.
The disadvantage of the
latter approach is an ambiguity in the choice of the multipole
operator which excites a particular mode, especially for
multicomponent systems\cite{bfsumrule}. Here we develop an
efficient numerical procedure for solving the hydrodynamical equations
in 1D, which doesn't involve additional ``sum rule''
approximation. After linearization, a system of hydrodynamic
equations can be written as
\begin{widetext}
\begin{eqnarray}
- m \omega ^2 \left[\begin{array}{c}
\delta \mu_b(x) \\
\delta \mu_f(x)
\end{array}\right] =
\left[\begin{array}{cc}
  \frac{\partial\mu_b}{\partial{n_b}} & \frac{\partial\mu_b}{\partial{n_f}} \\
  \frac{\partial\mu_f}{\partial{n_b}} & \frac{\partial\mu_f}{\partial{n_f}}
  \end{array}\right]\nabla\left(\left[ \begin{array}{cc}
  n^0_b(x) & 0 \\
  0 & n^0_f(x)
  \end{array}\right]
  \nabla
\left[\begin{array}{c}
\delta \mu_b(x) \\
\delta \mu_f(x)
\end{array}\right]\right).
\label{eqmuinside}
\end{eqnarray}
\end{widetext}
In the outer fermionic shell, $\delta \mu^{out}_f$ satisfies an equation:
\begin{equation}
- m \omega ^2 \delta \mu^{out}_f=\frac{\partial \mu_f^{out}}{\partial{n_f}}\nabla\left[n^{out}_f(x)\nabla \delta \mu^{out}_f \right].
\label{eqout}
\end{equation}

All modes can be classified by their parity with respect to the
space inversion, $x\rightarrow-x.$ Thus, we need to solve
equations (\ref{eqmuinside}-\ref{eqout}) only in the region $x>0$. For even modes, one has
the following conditions at $x=0:$
\begin{eqnarray}
\nabla \delta \mu_f(x=0)=0, \, \nabla \delta \mu_b(x=0)=0.\label{evenbc}
\end{eqnarray}
For odd modes, analogous conditions are
\begin{eqnarray}
\delta \mu_f(x=0)=0, \,  \delta \mu_b(x=0)=0.\label{oddbc}
\end{eqnarray}

Boundary conditions for fermions at the edge of the bosonic cloud, $x_b,$ correspond to the continuity of $\delta \mu_f$ and $v_f:$
\begin{eqnarray}
\nabla \delta \mu^{out}_f (x=x_b+0)=\nabla \delta \mu_f(x=x_b-0),
\label{boundc1}\\
\delta \mu^{out}_f (x=x_b+0)= \delta \mu_f(x=x_b-0).
\label{boundc2}
\end{eqnarray}

Two additional conditions come from the absence of the bosonic(fermionic) flow at $x_b (x_f):$
\begin{eqnarray}
n^0_b(x) v_b(x)\vert_{x\rightarrow x_b-0}=0,\label{boseflow}\\
n^0_f(x) v_f(x)\vert_{x\rightarrow x_f-0}=0. \label{fermiflow}
\end{eqnarray}
Outside of the region of coexistence, the chemical potential and
density of fermions are given by  $\mu^{out}_f\sim (n^{out}_f)^2,
n^{out}_f \sim\sqrt{1-(x/x_f)^2},$ where $x_f$ is the fermionic
cloud size. For these functions, there exists a general nonzero
solution of eq. (\ref{eqout}) which satisfies (\ref{fermiflow}): 
$ \delta \mu_f^{out}= \cos((\omega/\omega_0)\arccos{x/x_f}).$

Substituting this into (\ref{boundc1})-(\ref{boundc2}), one has to
solve eigenmode equations numerically for $x<x_b,$ with five
boundary conditions
(\ref{boundc1}),(\ref{boundc2}),(\ref{boseflow}) and
(\ref{evenbc}) or (\ref{oddbc}) depending on the parity. These
boundary conditions are compatible, only if $\omega$ is an
eigenfrequency.  We choose to leave out condition (\ref{boseflow}), 
find a numerical solution to a system of differential equations,
and check later if (\ref{boseflow}) is satisfied to identify the eigenfrequencies.

\begin{figure}
\psfig{file=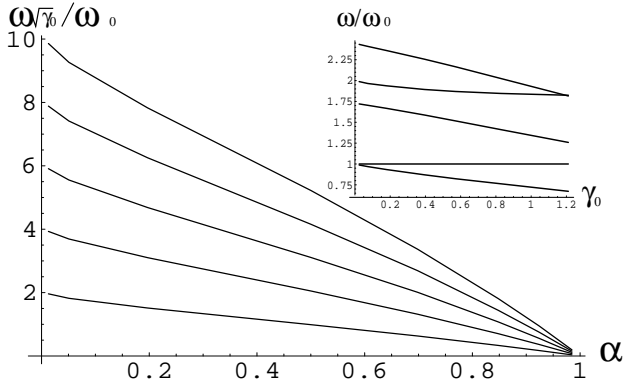}\caption{ \label{tgmodes} Dependence of the
energy of several lowest lying "out of phase" modes for $\gamma_0 \gg1$ on overall boson
fraction $\alpha,$ where $\gamma_0$ is the TG parameter in the
center of  a trap. Note, that we don't show "in phase" TG modes $\omega=n \omega_0,$ 
since their frequency is much higher. The inset shows frequencies of low-lying modes 
in mean-field regime for an equal overall number of bosons and fermions.}
\end{figure}

First we apply this numerical procedure for weakly-interacting 
regime, see the inset of fig. \ref{tgmodes}.
When $\gamma_0\rightarrow 0,$ bose and fermi clouds do not interact,
and collective modes coincide with purely bosonic or fermionic modes,
with frequencies\cite{Menotti} $\omega^f=n \omega_0$ and $\omega^b=\omega_0\sqrt{n(n+1)/2} .$ As interactions get stronger, bose and fermi clouds get coupled,
and all the modes except for dipole Kohn mode change their frequency.
For the case of strongly interacting  BF mixture,  we find
low-energy modes which correspond to the ``out of phase''
oscillations of bose and fermi atoms, and that keep the total
density approximately constant. These modes can be understood
as follows: for $\gamma \gg 1$ 
the energetic penalty for changing the relative
density of bosons and fermions is small. The frequency of such relative modes  scales as
 $\sim \omega_0/\sqrt{\gamma_0}.$ In fig. \ref{tgmodes} we show the numerically obtained dependence of the energy of these low energy modes as a function of the ratio
 of the total number of bosons and fermions.

\begin{figure}
\psfig{file=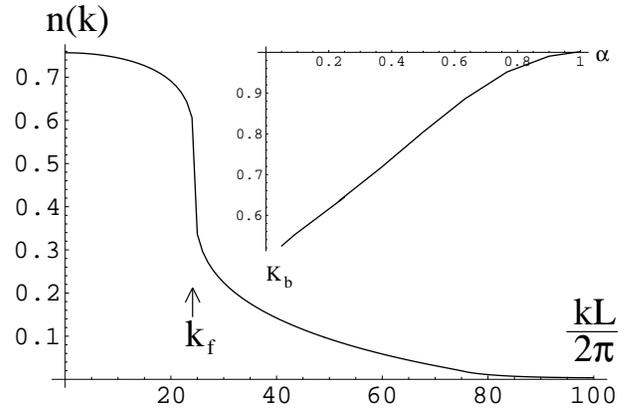} \caption{ \label{K} Fourier transform of fermi-fermi correlation function ($N_b=51,N=100$) - the fermi step gets strongly modified by interactions.  The inset shows bose-bose correlation LL parameter $K_b$ (see text)  as a function of bose fraction $\alpha$.
}
\end{figure}

Finally, we report the results of the calculations of
the bose-bose  and fermi-fermi correlation functions in systems
with periodic  boundary conditions. Such calculations can be
performed in the regime $\gamma \gg 1,$ due to ``factorization''
of the Bethe wavefunction, similar to \cite{OgataShiba}. 
"Spin" part of the wavefunction is related to the ground state 
of spin XY model, and significant analytical progress
can be made compared to the case of spin$-1/2$ fermion model \cite{OgataShiba}.
 Details of these calculations will be
presented elsewhere\cite{Bosefermilong}. The Fourier
transform of correlation function is  an occupation number $n(k),$
which can be measured directly in time-of-flight
experiments\cite{Paredes} or using Bragg spectroscopy\cite{Bragg}.
For fermions, $n(k)$ is shown in fig \ref{K}. The discontinuity at
$k_f=\pi n_f$ gets smeared out by interactions, but the overall
change in $n(k)$ after crossing $k_f$ remains quite large. For
bosons, $n(k)$ is a monotonous decreasing function, with a
singularity at $k=0.$ Singularity at $k=0$ is governed by the long
distance behavior of correlation function, characterized by the
bosonic LL\cite{Haldane, Cazalilla} parameter $K_b:$
\begin{equation}
\rho(0,\xi)\sim 1/|\xi|^{1/(2K_b)}, n(k)\sim |k|^{-1+1/(2K_b)},
\label{Keq}
\end{equation}
where $\xi$ is a distance between two points.
For periodic boundary conditions, one can extract $K_b$
using\cite{Cazalilla}  a fit $\rho(0,\xi)\sim 1/|\sin{(\pi
\xi/L)}|^{1/(2K_b)}$ for $\xi \gg L/N_b.$ Results of the
numerically extracted value of $K_b$  for strong interactions are
shown in fig \ref{K}: depending on the relative number of bosons
and fermions, $K_b$ changes from $0.5$ to $1.$ We note that in the
absence of fermions $K_b$  changes from $\infty$ to $1$ for
interacting bosons as the interactions get stronger. In the
presence of fermions, bosonic correlations decay faster than  in
the purely bosonic system. In principle, the low energy physics of
a BF mixture is characterized not only by bose-bose or fermi-fermi
correlations, but also by the long-range behavior of BF
correlations\cite{CazalillaHo, Mathey}. Our method  can be applied
to the evaluation of BF correlations and can be generalized to
nonzero temperatures.

We thank M. Lukin, L. Mathey, G. Shlyapnikov, D.Petrov, P.Wiegmann, C. Menotti and D.W. Wang for useful discussions. This work was partially supported by the NSF
grant DMR-0132874.


\begin{thebibliography}{10}
\bibitem{Feshbach}S. Inouye {\it et al.}, Phys. Rev. Lett. {\bf 93}, 183201 (2004);
F. Ferlaino {\it et al.}, cond-mat/0510630; C.A. Stan {\it et al.}, Phys. Rev. Lett. {\bf 93}, 143001 (2004).
\bibitem{Jaksch98}D. Jaksch {\it et al.}, Phys.Rev.Lett. {\bf 81}, 3108 (1998).
\bibitem{Bloch} M. Greiner {\it et al.}, Nature {\bf 415}, 39(2002).
\bibitem{Weiss} T. Kinoshita,  T. Wenger and D.S. Weiss,  Science, {\bf 305}, 1125 (2004).
\bibitem{Paredes} B. Paredes {\it et al.}, Nature {\bf 429}, 277 (2004).
\bibitem{Moritz1dmolecules} H. Moritz {\it et al.},  Phys. Rev. Lett. {\bf 94}, 210401 (2005).
\bibitem{bftheory}K. Molmer, Phys. Rev. Lett. {\bf 80}, 1804 (1998);
 L. Viverit, C. J. Pethick and  H. Smith, Phys. Rev.A {\bf 61}, 053605 (2000);
  M. Lewenstein {\it et al.}, Phys. Rev. Lett. {\bf 92}, 050401 (2004);
D.-W. Wang, M.Lukin and E.Demler, cond-mat/0410494.
\bibitem{bfexp}B. DeMarco and D.S. Jin, Science, {\bf 285}, 1703(1999);
 F. Schreck {\it et al.}, Phys. Rev. Lett. {\bf 87}, 080403 (2001);
G. Modugno {\it et al.}, Science  {\bf 297}, 2240 (2002); Z. Hadzibabic {\it et al.},
 Phys. Rev. Lett. {\bf 88}, 160401 (2002); J. Goldwin {\it et al.},  Phys. Rev. A {\bf 70}, 021601(R) (2004).
\bibitem{fermipressure}A.G. Truscott {\it et al.}, Science {\bf 291}, 2570(2001).
\bibitem{Giamarchi}T. Giamarchi, {\it Quantum physics in One dimension}, Clarendon Press (Oxford, UK, 2004).
\bibitem{Das}K.K. Das,  Phys. Rev. Lett. {\bf 90}, 170403 (2003).
\bibitem{CazalillaHo}M. A. Cazalilla and  A. F. Ho, Phys. Rev. Lett. {\bf 91}, 150403 (2003).
\bibitem{CDW}T. Miyakawa, H. Yabu and  T. Suzuki, Phys. Rev. A {\bf 70}, 013612 (2004);
E. Nakano and H.Yabu,  Phys. Rev. A {\bf 72}, 043602 (2005).
\bibitem{Mathey} L. Mathey {\it et al.}, Phys. Rev. Lett. {\bf 93}, 120404 (2004).
\bibitem{Moritz}H. Moritz {\it et al.}, Phys. Rev. Lett. {\bf 91}, 250402 (2003).
\bibitem{BECinbox} T.P. Meyrath {\it et al.},  Phys. Rev. A, 71, 041604(R) (2005).
\bibitem{Cote} R. Cote {\it et al.}, Phys. Rev. A {\bf 57},R4118(1998) .
\bibitem{Yb} K. Honda {\it et al.},  Phys. Rev. A {\bf 66}, 021401(R) (2002);
 Y. Takasu {\it et al.}, Phys. Rev. Lett. {\bf91}, 040404 (2003);
 C. Y. Park and T. H. Yoon , Phys. Rev. A {\bf 68}, 055401 (2003).
\bibitem{Rb} J.P. Burke and J.L. Bohn, Phys. Rev. A {\bf 59},1303(1999);
S. G. Crane {\it et al.}, Phys. Rev. A{\bf  62}, 011402(R) (2000).
\bibitem{Olshanii98} M.Olshanii, Phys. Rev. Lett. {\bf 81}, 938(1998).
\bibitem{Lai}C.K. Lai and C.N.Yang, Phys. Rev A, {\bf 3}, 393(1971);
C.K.Lai Journ. of Math. Phys., 15, 954(1974).
\bibitem{LL}E.H. Lieb and W. Liniger, Phys. Rev. {\bf 130}, 1605 (1963).
\bibitem{Menotti}C. Menotti and S. Stringari, Phys. Rev. A {\bf 66}, 043610 (2002).
\bibitem{Stringari}S. Stringari,  Phys. Rev. Lett. {\bf 77}, 2360 (1996).
\bibitem{Astrakharchik} G.E.  Astrakharchik {\it et al}, Phys. Rev. Lett. {\bf 93}, 050402 (2004).
\bibitem{bfsumrule}T. Miyakawa, T. Suzuki and H. Yabu, Phys. Rev. A {\bf62}, 063613 (2000).
\bibitem{OgataShiba} M. Ogata and H. Shiba, Phys. Rev. B {\bf 41}, 2326 (1990).
\bibitem{Bosefermilong}A. Imambekov and E.Demler, cond-mat/0510801.
\bibitem{Bragg}S. Richard {\it et al.}, Phys. Rev. Lett. {\bf 91}, 010405 (2003).
\bibitem{Haldane} F.D.M. Haldane, Phys.Rev. Lett. {\bf 47}, 1840 (1981).
\bibitem{Cazalilla}M. A. Cazalilla, Journal of Physics B:  AMOP 37, S1-S47 (2004).
\end{thebibliography}
\end{document}